\documentclass[aps,prl,twocolumn,superscriptaddress]{revtex4}

\usepackage{mathrsfs,amssymb,graphics,subfigure,threeparttable}
\usepackage{grap hicx}
\usepackage{amssymb}
\usepackage{enumerate}
\usepackage{amsmath}
\usepackage{fancyhdr}
\usepackage{bm}
\usepackage[squaren]{SIunits}
\usepackage{hyperref}

\begin{document}

% Use the \preprint command to place your local institutional report
% number in the upper righthand corner of the title page in preprint mode.
% Multiple \preprint commands are allowed.
% Use the 'preprintnumbers' class option to override journal defaults
% to display numbers if necessary
%\preprint{}

%Title of paper
\title{Phase space dynamics of a plasma wakefield dechirper for energy spread reduction}

\author{Y. P. Wu}
\affiliation{Department of Engineering Physics, Tsinghua University, Beijing 100084, China}
\author{J. F. Hua}
\email[]{jfhua@tsinghua.edu.cn}
\affiliation{Department of Engineering Physics, Tsinghua University, Beijing 100084, China}
\author{Z. Zhou}
\affiliation{Department of Engineering Physics, Tsinghua University, Beijing 100084, China}
\author{J. Zhang}
\affiliation{Department of Engineering Physics, Tsinghua University, Beijing 100084, China}
\author{S. Liu}
\affiliation{Department of Engineering Physics, Tsinghua University, Beijing 100084, China}
\author{B. Peng}
\affiliation{Department of Engineering Physics, Tsinghua University, Beijing 100084, China}
\author{Y. Fang}
\affiliation{Department of Engineering Physics, Tsinghua University, Beijing 100084, China}
\author{Z. Nie}
\affiliation{Department of Engineering Physics, Tsinghua University, Beijing 100084, China}
\author{X. N. Ning}
\affiliation{Department of Engineering Physics, Tsinghua University, Beijing 100084, China}
\author{C. H. Pai}
\affiliation{Department of Engineering Physics, Tsinghua University, Beijing 100084, China}
\author{Y. C. Du}
\affiliation{Department of Engineering Physics, Tsinghua University, Beijing 100084, China}
\author{W. Lu}
\email[]{weilu@tsinghua.edu.cn}
\affiliation{Department of Engineering Physics, Tsinghua University, Beijing 100084, China}
\author{C. J. Zhang}
\affiliation{University of Los Angeles, Los Angeles, California 90095, USA}
\author{W. B. Mori}
\affiliation{University of Los Angeles, Los Angeles, California 90095, USA}
\author{C. Joshi}
\affiliation{University of Los Angeles, Los Angeles, California 90095, USA}

\date{\today}

\begin{abstract}
Plasma-based accelerators have made impressive progress in recent years. However, the beam energy spread obtained in these accelerators is still at $\sim 1\%$ level, nearly one order of magnitude larger than 
what is needed for challenging applications like coherent light sources or colliders.
In plasma accelerators, the beam energy spread is mainly dominated by its energy chirp (longitudinally correlated energy spread).
Here we demonstrate that when an initially chirped electron beam from a linac with a proper current profile is sent through a low-density plasma structure, the self wake of the beam can significantly reduce its energy chirp and the overall energy spread. 
The resolution-limited energy spectrum measurements show at least a threefold reduction of the beam energy spread from 1.28 $\%$ to 0.41 $\%$ FWHM with a dechirping strength of $\sim$ 1 (MV/m)/(mm pC).
Refined time-resolved phase space measurements, combined with high-fidelity three-dimensional particle-in-cell simulations, further indicate the real energy spread after the dechirper is only about $0.13\ \%$ (FWHM), a factor of 10 reduction of the initial energy spread.

\end{abstract}


\maketitle

In recent years, great strides have been made in the field of plasma-based wakefield accelerators \cite{Nature_2004_1, Nature_2004_2, Nature_2004_3,leemans_4gev, blumenfeld2007energy, Nature_high_efficiency,Nature_positron,NC_positron_driven_hollow_channel_PWFA}. 
However, the energy spread in these accelerators is typically at a few percent level, which is still nearly one order of magnitude larger than that is required in forefront applications like free electron lasers and linear colliders.
The beam energy spread in plasma accelerators is usually dominated by a nearly linear energy chirp arising from the relatively broad acceleration phase occupied by the beam. 
Therefore, reducing the beam energy spread down to $\sim 0.1$ percent level requires effective energy chirp reduction in plasma accelerators.

Energy dechirpers based on corrugated wall structures or dielectric-based slab structures have been experimentally demonstrated \cite{emma2014experimental, Antipov2014experimental}.
To date, the typical dechirping strength $S_d$ experimentally obtained in these devices with mm-level gap 
is $\sim$ 0.01-0.1 (MV/m)/(mm pC) for $\sim$ ps-long electron beams, where $S_d$ is defined as the dechirping field divided by the bunch length and the bunch charge \cite{Antipov2014experimental}.  
The beam produced by a plasma accelerator is typically very short (few fs), and has a relatively large energy chirp (few to 10s MeV).
Therefore, it may be impractical to adopt a dechirper with such low $S_d$. 
Although theoretical studies indicate that $S_d$ can be improved to $\sim$ 10$^3$ (MV/m)/(mm pC) for 10 fs electron beams by reducing the gap size to 10s $\mu$m level \cite{Dielectric_dechirper}, such small gap may prove to be challenging for beam alignment.

\begin{figure}[tp]
\includegraphics[height=0.165\textwidth]{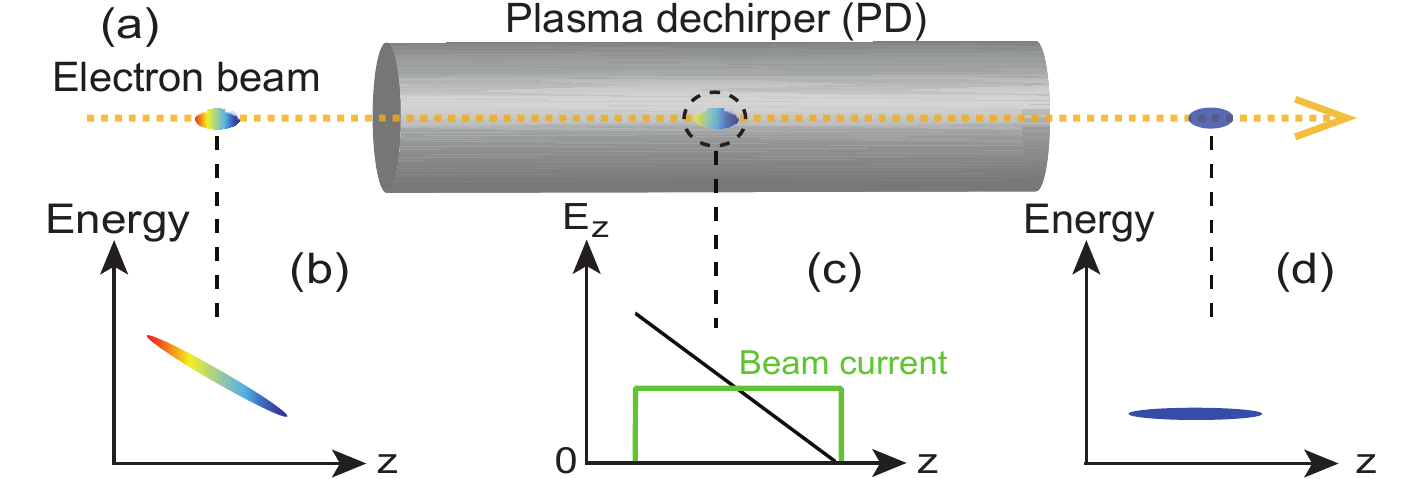}
\caption{\label{fig1}
(a) Schematic diagram of the PD.
(b) The beam longitudinal phasespace at the entrance of the PD. (c) The beam current profile (green line) and the on-axis longitudinal wakefield $E_z$ excited by the beam in the PD (black line).
(d) The final beam longitudinal phasespace.}
\end{figure}

In order to achieve a higher dechirping strength in practice, an alternative approach is to use a tunable plasma dechirper (PD) \cite{dechirper_IPAC_talk, dechirper_IPAC_paper2}, as shown schematically in Fig. 1(a). In this scheme, an electron beam with a nearly linear positive energy chirp (Fig. 1(b), the beam energy increases quasi-linearly from head to tail, which is normal for an underloaded wake in a plasma accelerator \cite{LifeiInjection, NC_chirp_compensation}) is sent through a separate low-density plasma section to excite a nearly linear plasma wake \cite{LimitsOfLinearPlasmaWakefieldTheory}.
For bunch lengths much shorter than the plasma wavelength, the beam will totally stay in the decelerating phase of the wake where $E_z$ has a negative slope (Fig. 1(c)), where the beam tail experiences a larger energy loss than the head. 
Therefore, the positive chirp can be effectively eliminated during the propagation (Fig. 1(d)).
The total dechirping effects can be easily tuned by changing the density and length of the plasma.

The concept of a PD was first proposeded and experimentally demonstrated by Tsinghua group in 2017, where an energy chirp reduction of factor 1.25 was clearly observed \cite{dechirper_IPAC_talk, dechirper_IPAC_paper2}. Very recently, this method has been extended to show a factor 4-6 reduction of the energy chirp through energy spectrum measurement \cite{FlashForward_dechirper,INFN_dechirper}. However, for this method to be really useful for reducing energy spread of electron beams in plasma accelerators down to 0.1 $\%$ level, a factor 10 reduction of the energy chirp and overall energy spread is needed, 
where the complex interplay and trade-off among the linear chirp reduction, the nonlinear chirp increase and the slice energy spread growth become critical.
In this Letter, we demonstrate a near tenfold beam energy spread reduction in a properly designed beam current profile and PD parameters, and the complex longitudinal phase space dynamics of the dechirping process is clearly revealed through refined time-resolved phase space measurements, in good agreement with high-fidelity three-dimensional (3D) particle-in-cell (PIC) simulations.

To quantify the effects of a PD, we show in Fig. 2 simulation results obtained by full 3D PIC code QuickPIC \cite{QuickPIC_1, QuickPIC_2} for three typical beam current profiles (flat-top, parabolic and Gaussian, Fig. 2 (a)). 
In these simulations, electron beams with a transverse Gaussian profile ($\sigma_r=0.15\ k_p^{-1}$) are initialized with zero slice energy spread and positive linear chirp (FWHM energy spread $\Delta W_i$), and the plasma is initialized with a uniform electron density $n_p$, where $k_p^{-1}$ is the plasma skin depth.

\begin{figure}[tp]
\includegraphics[height=0.345\textwidth]{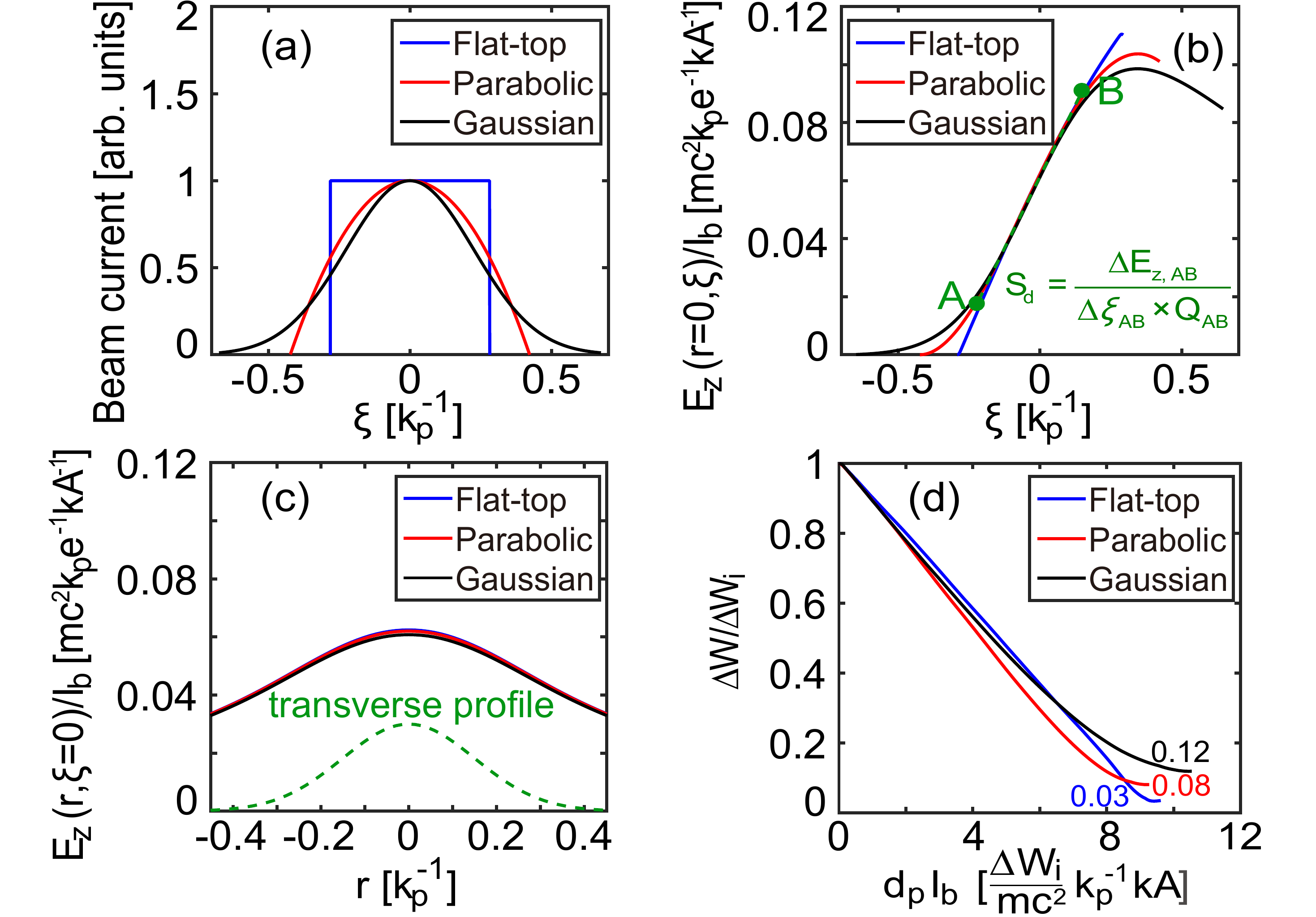}
\caption{\label{fig1} (a) Three current profiles for electron beams with the same total charge $Q$, peak density $n_b$ and transverse size $\sigma_r$, here $\xi=ct-z$ and $\xi<0$ corresponds to the beam head. 
(b) and (c) show $E_z$ ($r=0$)/$I_b$ versus $\xi$ and $E_z$ ($\xi=0$)/$I_b$ versus $r$ for these three current profiles, respectively.
Here $E_z$ is normalized to the plasma wave-breaking limit $mc^2k_p/e$ and $I_b=2\pi \sigma_r^2 n_bec\approx 0.2(n_b/n_p)$ kA. The dechirping strength $S_d$ is calculated using the linear fit to $E_z$ between points A and B. (d) Evolution of the beam FWHM energy spread $\Delta W$ (normalized to $\Delta W_i$) versus $d_pI_b$ for these three current profiles.
}
\end{figure}

In Fig. 2(b), the on-axis $E_z$s (divided by the beam peak current $I_b$) are plotted, where a similar linear slope near the beam center is obtained for all three profiles. 
For the flat-top profile, $E_z$ is close to linear within the whole beam, while for non flat-top profiles, $E_z$s have nonlinear forms near the head and the tail, which can induce nonlinear energy chirps during the dechirping process.
In Fig. 2(c), the radial dependences of $E_z$s are plotted to show the transverse non-uniformity of the wake, which can induce beam slice energy spread growth during the dechirping process. 
The final achievable minimum energy spread of the beam is determined by a trade-off among the linear chirp reduction, the nonlinear chirp increase and the slice energy spread growth.
In Fig. 2(d), the energy spread (FWHM) reduction versus the product of the propagation distance $d_p$ and the peak current $I_b$ are plotted, where $d_p$ is normalized to $\frac{\Delta W_i}{mc^2}k_p^{-1}$.
It can be clearly seen that after a distance of about 9.5 (flat-top)/9.2 (parabolic)/10.5 (Gaussian) $\frac{\Delta W_i}{mc^2}(k_pI_b$[kA])$^{-1}$, the minimum energy spread about 3$\%$ (flat-top)/8 $\%$ (parabolic)/12 $\%$ (Gaussian) of its initial value can be achieved, which suggests a dechirping factor of  33.3 (flat-top)/12.5 (parabolic)/8.3 (Gaussian).
Based on Fig. 2(b), the dechirping strength can be estimated as $S_d \approx 3.2\times 10^5 (n_p[$cm$^{-3}]/10^{18})^{3/2}$ (MV/m)/(mm pC). 
For a given beam, the plasma wavelength must be larger than the bunch length. 
For example, for bunch length $\sim k_p^{-1}$, plasma densities vary from $\sim 5\times10^{18}$ cm$^{-3}$ to $5\times 10^{14}$ cm$^{-3}$ for 10 fs$-$1 ps beams, giving a $S_d$ $\sim 10^6- 1$ (MV/m)/(mm pC).

To confirm the above predictions, we have performed a plasma dechirping experiment on the TTX platform at Tsinghua University \cite{TTX2, 10-40MeV@THU}.
The schematic layout is shown in Fig. 3(a). 
A 40 pC, 1.1 ps (FWHM), 46 MeV electron beam with a positive linear energy chirp is generated by a high brightness S-band RF linac.
The bunch charge is set by tuning the energy of the 300 fs (FWHM), 266 nm photocathode drive laser.
The bunch length is achieved through velocity compression in this photogun by launching the beam at low phase \cite{compression_effect_in_gun}.
The positive energy chirp is imprinted by off-crest acceleration in the accelerating tube.
The electron beam is focused to a transverse size $\sigma_r = 40\ \mu$m (Fig. 3(b)) at the 
front edge of a slit gas jet by two triplets, and detected by a removable OTR screen (Screen1 in Fig. 3(a)). 
The normalized emittance of the beam is measured to be $\sim 1.5$ mm mrad by using a two-screen method (Screen1 and Screen2 about 2 m downstream) \cite{emittance_measurement_two_screen_method}.
The beam longitudinal phase space is measured on another YAG:Ce screen (Screen3 in Fig. 3(a)) by using a RF deflecting cavity (temporal resolution $\sim$ 0.4 ps FWHM) and a dipole magnet, as shown in Fig. 3(c). By combining the longitudinal phase space measurement with beam vertical distribution (deflecting cavity off), the beam current profile can be obtained through deconvolution, as shown in Fig. 3(d).
We note that this profile is similar to a parabolic distribution, and a plasma density $\lesssim$ $5\times 10^{14}$ cm$^{-3}$ should be used to ensure the monotonicity of dechirping field within the beam.

\begin{figure*}[tp]
\includegraphics[height=0.36\textwidth]{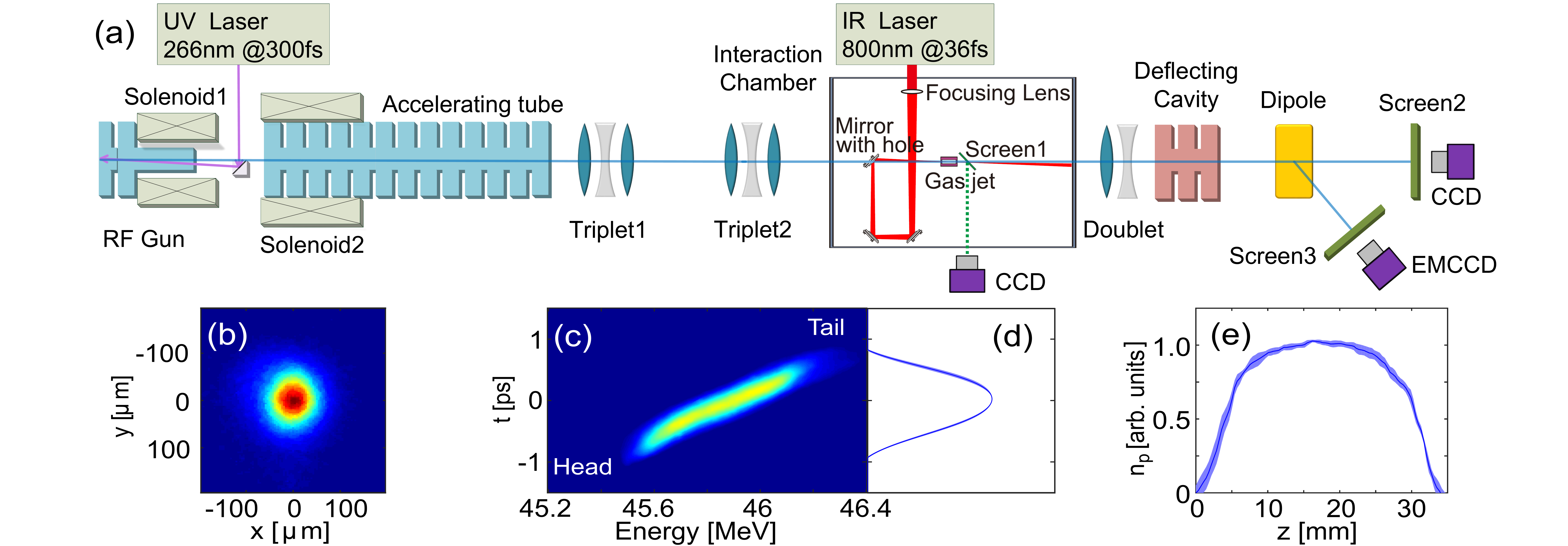}
\caption{\label{fig2} (a) Schematic layout of the plasma dechirping experiment. (b) The beam waist profile. (c) Measured beam longitudinal phase space. (d) Deconvoluted beam current profile. The shaded regions correspond to the standard deviation of 20 consecutive shots. (e) Longitudinal distribution of plasma density through off-line measurement, where the shaded regions correspond to the standard deviation of 10 consecutive shots.}
\end{figure*}

To generate a low-density plasma with $n_p\lesssim$ $5\times 10^{14}$ cm$^{-3}$, a method based on laser ionization of a mixed gas ($1\%$ H$_2$ + $99\%$ He) is used, where the laser intensity is chosen properly to just ionize the hydrogen atoms.
The longitudinal gas profile from the slit gas jet (30 mm by 2 mm) was measured off line using shearing interferometry by a wavefront sensor \cite{wavefront_sensor} with Argon gas, as shown in Fig. 3 (d).
A $36$ fs (FWHM) 800 nm laser pulse is focused to a waist size $w_0 \sim 110\ \mu$m by a lens ($f=1500$ mm) near the center of the slit gas jet. Right after ionization occurs, the plasma approximately has an initial radius $\sim w_0$ and a density in the range $10^{15}-10^{16}$ cm$^{-3}$, which is proportional to the backing gas pressure $P_g$ ($0.5-5$ MPa).
After a proper delay ($\sim$ 10 ns), the plasma expands to a wider size with a lower density approximately in the range of $10^{14}-10^{15}$ cm$^{-3}$.
As shown in Ref. \cite{plasma_expansion}, the plasma expansion rate is dominated by the initial electron temperature induced by the ionization process, and has little dependence on the initial density.
As a result, the plasma density after expansion with given delay is approximately proportional to its initial backing pressure $P_g$.

To demonstrate the dechirping effect with this low-density plasma, the electron beam was sent through a $3$-mm central hole on the final turning mirror of the laser pulse to focus right near the front edge of the gas jet.
The laser pulse collinearly propagates with and arrives about $10$ ns before the electron beam with a timing jitter of $\sim$ 100 fs \cite{synchronization_TTX}. The electron beam has a negligible transverse beam position jitter at the focus ($\sim$ 3 $\mu$m), and propagates through the $\sim$ 30 mm-long low-density plasma. 
Figure 4(a) shows the energy spectrum of the incoming beam on Screen3 after dispersion by the dipole magnet when the plasma is off. 
Integrated energy spectra of 20 consecutive shots similar to Fig. 4(a) are shown in Fig. 4(e) (the blue band), which gives a FWHM energy spread of 0.59 MeV.
Figure 4(b), (c) and (d) show the spectrum of the outcoming beam on Screen3 for three different backing pressures (0.5, 1 and 2 MPa), and the effect of dechirping is evident. The integrated energy spectra are also shown in Fig. 4(e) with different colors, as the gas pressure is increased the energy spread becomes smaller. 
For $P_g=2$ MPa, the FWHM energy spread from Fig. 4(e) (the red line) is $0.19$ MeV, which gives a more than threefold reduction in the absolute energy spread, and a reduction in the relative energy spread from $1.28 \%$ to $0.41\%$. 
In these measurements, the horizontal size and the divergence of the beam at the entrance of the dipole limit the energy resolution to about 0.17 MeV (FWHM), and this can be estimated directly from the expanded slice energy spread obtained by longitudinal phase space measurement in Fig. 3(c), where the true slice energy spread is below 0.01 MeV based on simulations of our beamline and measurements of similar beamlines \cite{LCLS_laser_heater, measurement_local_energy_spread}.
Therefore, the threefold reduction in energy spread should be considered only as a lower limit.

\begin{figure}[tp]
\includegraphics[height=0.196\textwidth]{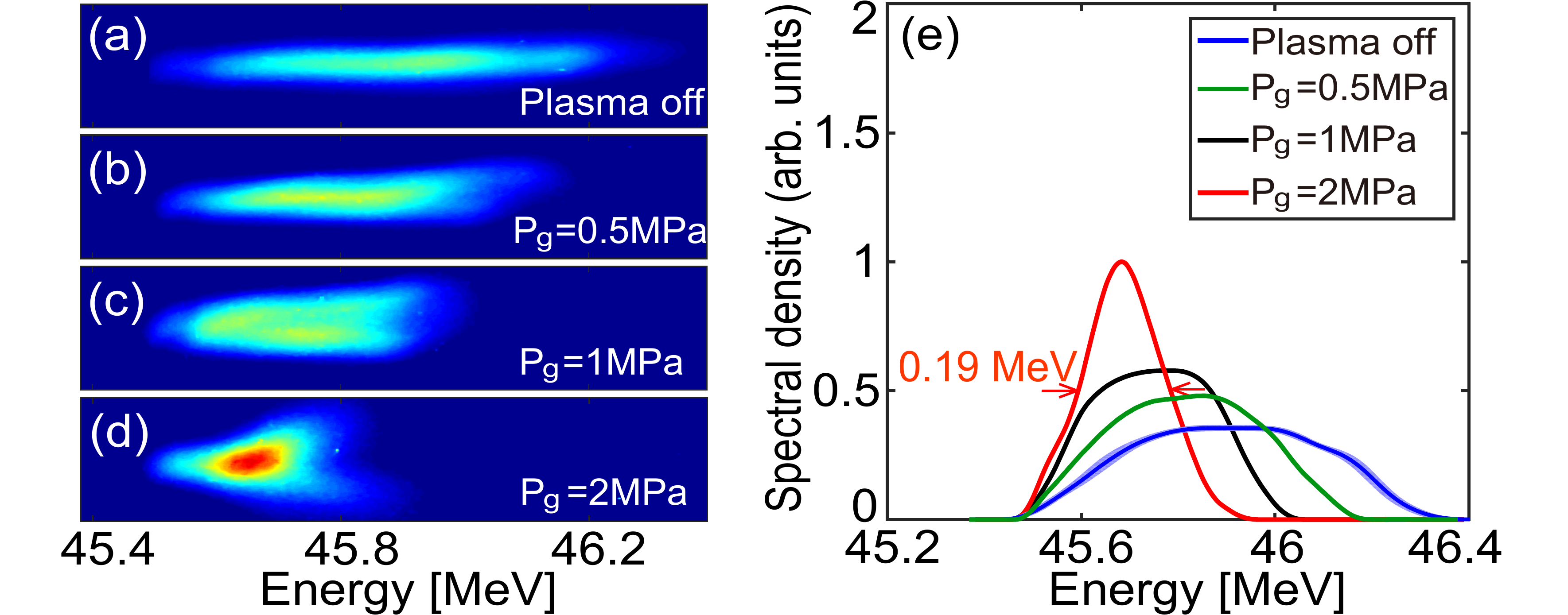}
\caption{\label{fig7} (a)-(d) Beam energy spectrum distributions on Screen3 with plasma off, $P_g=0.5$ MPa, $P_g=1$ MPa and $P_g=2$ MPa, respectively. (e) Beam energy spectra derived from the images in the left panel. For the plasma-off case, the shaded regions correspond to the standard deviation of 20 consecutive shots. 
}
\end{figure}

To get a deeper insight of the dechirping process, and also to alleviate the effect of the limited energy resolution,
we performed refined beam longitudinal phase space measurements, and made detailed comparisons with high-fidelity 3D PIC simulations using the code QuickPIC.
Three measured beam longitudinal phase spaces on Screen3 after dechirping with $P_g=0.5$ MPa, $P_g=1$ MPa and $P_g=2$ MPa are shown in Fig. 5(b), (c) and (d), and the corresponding projected energy spectra are shown in Fig. 5(m) with green, black and red solid lines, where the projected FWHM energy spread for $P_g=2$ MPa is $0.19$ MeV, very similar to the result obtained from the direct energy spectrum measurement (Fig. 4(e)).

To make comparisons between the high-fidelity simulations and the experimental results, we use beam and plasma parameters in the simulations as close to the experimental conditions as possible. For the beam parameters, the measured current profile (Fig. 3(d)) and the energy chirp deduced from the centroid of the longitudinal phase space (purple line in Fig. 5(a)) are used, and the beam slice energy spread is also set to the upper limit (0.01 MeV FWHM) \cite{LCLS_laser_heater, measurement_local_energy_spread}. 
For the plasma parameters, the longitudinal plasma profile is set as the measured distribution in Fig. 3(e), and the plasma density $n_p$ is assumed to be proportional to the backing pressure $P_g$ as discussed before.
In Fig. 5(f) through (h), we show the simulated phase spaces on Screen3 obtained by scanning the single free parameter $n_p$ to get a best fit to the experimental measurements ($n_p=1\times 10^{14}$ cm$^{-3}$ for $P_g=0.5$ MPa, therefore $n_p=2\times 10^{14}$ cm$^{-3}$/$n_p=4\times 10^{14}$ cm$^{-3}$ for $P_g=1$ MPa/$P_g=2$ MPa as discussed before).
Here the effect of beam transport through the beamline downstream the plasma is fully taken into account.
In Fig. 5(m) we show a direct comparison between the measured (solid) and the simulated (dotted) integrated energy spectra. The agreement between the two is excellent for all three values of $P_g$. 
The above comparisons use only one parameter to closely match three longitudinal phase spaces and integrated energy spectra, 
giving us confidence for the value of the plasma density used, which is too low to be directly measured online by interferometry.

\begin{figure}[bp]
\includegraphics[height=0.645\textwidth]{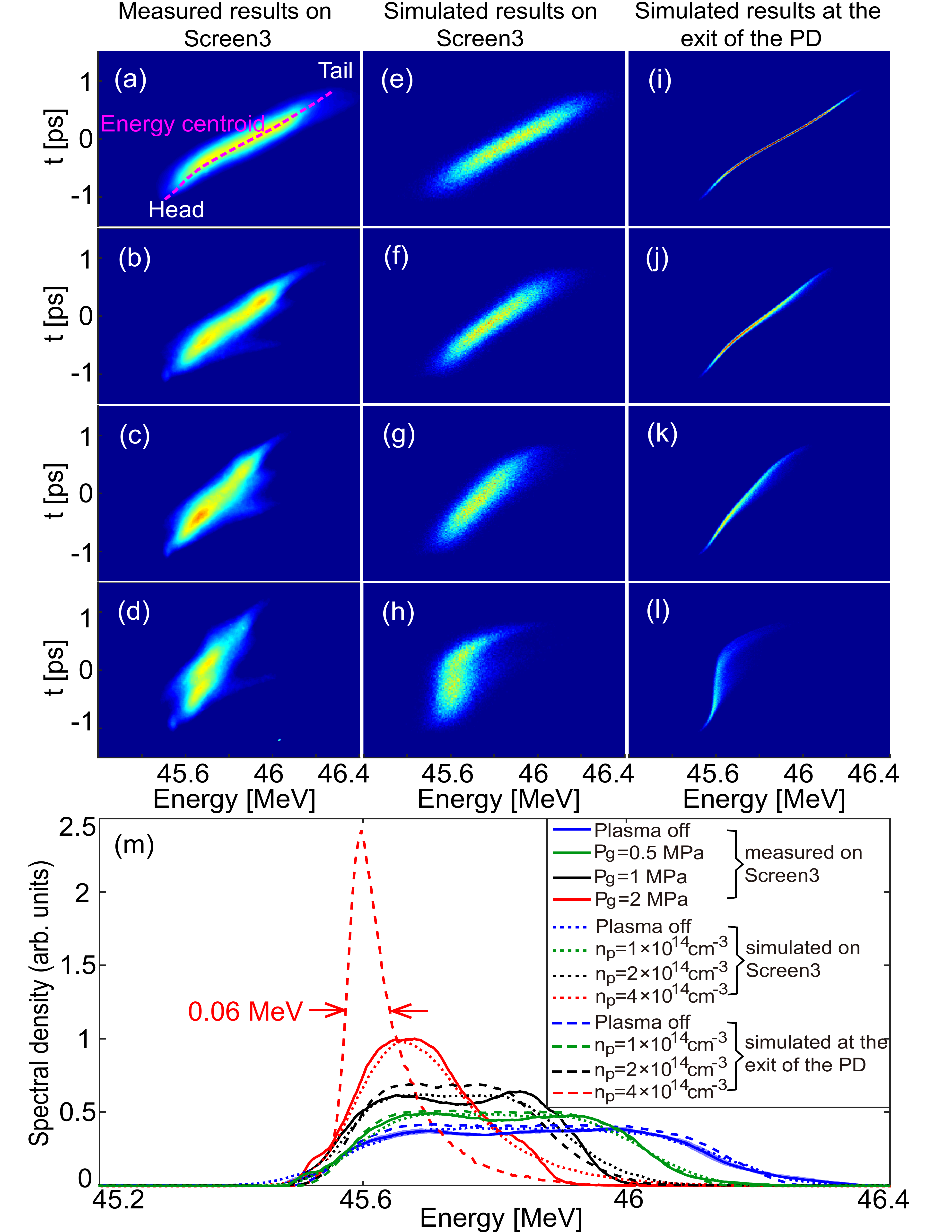}
\caption{\label{fig4} 
(a)-(l) Beam longitudinal phase space distributions (left/middle column: the experimental/simulated results recorded on Screen3; right column: the simulated results at the exit of the PD). The top row of images is for plasma off, the second, third and bottom rows are for plasma on with $P_g=0.5$ MPa/$n_p=1\times 10^{14}$ cm$^{-3}$, $P_g=1$ MPa/$n_p=2\times 10^{14}$ cm$^{-3}$ and $P_g=2$ MPa/$n_p=4\times 10^{14}$ cm$^{-3}$, respectively.
(m) The energy spectra of above cases. For the measured plasma-off case (solid blue line), the shaded regions correspond to the standard deviation of 20 consecutive shots.
}
\end{figure}

Based on the good agreements above, we can also get valuable information on the exact beam energy spread after the PD, which can not be directly measured due to the limited energy resolution about 0.17 MeV.
Figure 5(i) through (l) plot the simulated longitudinal phase spaces just at the exit of the PD for the cases of plasma off, 
$n_p=1\times 10^{14}$ cm$^{-3}$, $n_p=2\times 10^{14}$ cm$^{-3}$ and $n_p=4\times 10^{14}$ cm$^{-3}$, respectively. 
One can see that during the whole dechirping process, the slice energy spread increases slightly from its initial value (0.01 MeV) to less than 0.04 MeV (0.02 MeV for $n_p=1\times 10^{14}$ cm$^{-3}$, 0.03 MeV for $n_p=2\times 10^{14}$ cm$^{-3}$ and 0.04 MeV for $n_p=4\times 10^{14}$ cm$^{-3}$). 
The integrated energy spectra for these three cases are also shown in Fig. 5(m) with green, black and red dashed lines.
For $n_p=1\times 10^{14}$ cm$^{-3}$/$n_p=2\times 10^{14}$ cm$^{-3}$, the three green/black lines (solid for measured on Screen3, dotted for simulated on Screen3 and dashed for simulated at the exit of the PD) match each other well, suggesting that at this density the energy spread is still dominated by the residual energy chirp.
However, for $n_p=4\times 10^{14}$ cm$^{-3}$, the dashed red line (simulated energy spectrum at the exit of the PD) shows dramatically narrower spread compared with the other two lower density cases, which strongly suggests that at this density the real energy spread of the beam is much smaller (about $0.06$ MeV) than the resolution-limited measurement (0.19 MeV).
Therefore, the above simulations and comparisons with experimental phase space measurements suggest that the FWHM energy spread of the beam has been reduced from $0.59$ MeV to $0.06$ MeV for the case of $n_p=4\times 10^{14}$ cm$^{-3}$, leading to a near tenfold reduction in the relative energy spread (from 1.28 $\%$ down to 0.13 $\%$). 
This is also consistent with the simulation predictions of the final achievable minimum energy spread shown in Fig. 2(d) when a parabolic current profile is used. 
Furthermore, the simulated $E_z$ field in this case has a maximum strength of $\sim 15$ MV/m. Combining with the bunch charge and length in our experiment, this yields a dechirping strength of $S_d \approx 1$ (MV/m)/(mm pC), one to two orders of magnitude higher than those obtained using corrugated wall devices or dielectric-based slab structures as dechirpers for similar beams \cite{emma2014experimental, Antipov2014experimental}.

In this experiment, to resolve the beam longitudinal phase space with a resolution-limited deflecting cavity, a relatively long bunch length ($\sim$ 1.1 ps FWHM) is used.
These experimental results can be properly scaled for much shorter beams typically obtained in a plasma accelerator.
For example, for a 10 fs electron beam, roughly ten thousand times higher plasma density should be used ($n_p=4\times 10^{18}$ cm$^{-3}$) for the PD, and the corresponding dechirping strength is increased by a factor of $10^6$ according to previous analyses.

In summary, a PD scheme based on the beam's self-generated linear wake in a low-density plasma is proposed and experimentally demonstrated.
The experimental results, combined with high-fidelity 3D PIC simulations indicate a near tenfold reduction of the beam energy spread from 1.28 $\%$ to 0.13 $\%$ FWHM.
This tunable and flexible technique can be applied to future plasma-based photon sources and colliders for significantly enhancing the beam quality.

\section{ACKNOWLEDGMENTS}
The authors thank Dr. B. Liu for the support of quadrupole magnets and Dr. W. J. Ma for the support of diamond films for vacuum isolation. This work is supported by the NSFC Grants (No. 11535006, No. 11425521, No. 11775125 and No. 11875175), CAS Center for Excellence in Particle Physics, and the U.S. Department of Energy Grants (No. DE-SC0008491, No. DE-SC0008316 and DE-SC0010064) at UCLA. The numerical simulations were carried out using Sunway TaihuLight cluster at National Supercomputing Center in Wuxi (NSCCWX).

\section{references}

\end{document}